\documentclass[ALICE,manyauthors]{cernphprep}
\pdfoutput=1 
\usepackage[comma,square,numbers,sort&compress]{natbib}
\usepackage{hyperref}
\usepackage{graphicx}  
\usepackage{dcolumn}   
\usepackage{bm}        
\usepackage{amssymb}   
\usepackage{amsfonts}
\usepackage{graphics}
\usepackage{epsfig}
\usepackage[normalem]{ulem}
\usepackage[usenames]{color}
\usepackage{multirow}

\newcommand{\VZERO}        {\rm{VZERO}}

\newcommand{\pbar}         {$\rm\overline{p}$}

\newcommand{\dedx}         {\ensuremath{\mathrm{d}E/\mathrm{d}x}}

\newcommand{\PbPb}         {\mbox{Pb--Pb}}
\newcommand{\AuAu}         {\mbox{Au--Au}}

\newcommand{\dNdeta}       {\ensuremath{\mathrm{d}N_\mathrm{ch}/\mathrm{d}\eta}}

\newcommand{\dcaxy}        {\ensuremath{{\rm DCA}_{xy}}}
\newcommand{\dcaz}         {\ensuremath{{\rm DCA}_{z}}}
\newcommand{\EcrossB}      {E$\times$B}

\newcommand{\pt}           {\ensuremath{p_{\rm T}}}
\newcommand{\ppi}          {\ensuremath{{\rm p}/\pi}}
\newcommand{\kpi}          {\ensuremath{{\rm K}/\pi}}

\newcommand{\snn}          {\ensuremath{\sqrt{s_{\rm NN}}}}

\newcommand{\Tfo}          {\ensuremath{{T}_{kin}}}
\newcommand{\Tch}          {\ensuremath{{T}_{ch}}}

\newcommand{\avbT}         {\ensuremath{\left< \beta_{\rm T}\right>}}

\newcommand{\muB}          {\ensuremath{\mu_{B}}}

\newcommand{\gevc}[1]      {\ensuremath{{\rm GeV}/c}}

\newcommand{\warn}[1]      {}

\graphicspath{{./img/}}
\renewcommand{\warn}[1]{} 

\begin{document}%
%
%
\begin{titlepage}
\PHnumber{2012-230}                 
\PHdate{08 Aug 2012}              
%
%
\title{Pion, kaon, and proton production\\in central \PbPb\ collisions at \snn~=~2.76~TeV}
\ShortTitle{$\pi$, K, and p production in central \PbPb\ collisions at \snn~=~2.76~TeV}   
%
\Collaboration{ALICE Collaboration%
         \thanks{See Appendix~\ref{app:collab} for the list of collaboration members}}
\ShortAuthor{ALICE Collaboration}      
\begin{abstract}
  In this Letter we report the first results on $\pi^\pm$, K$^\pm$, p and \pbar\
  production at mid-rapidity ($\left|y\right|<0.5$) in central \PbPb\
  collisions at \snn~=~2.76~TeV, measured by the ALICE experiment at
  the LHC. The \pt\ distributions and yields are compared to
  previous results at \snn~=~200~GeV and expectations from 
  hydrodynamic and thermal models.  The spectral shapes indicate a
  strong increase of the radial flow velocity with \snn, which in
  hydrodynamic models is expected as a consequence of the increasing
  particle density. While the \kpi\ ratio is in line with predictions
  from the thermal model, the \ppi\ ratio is found to be lower by a
  factor of about 1.5. This deviation from thermal model
  expectations is still to be understood.


\end{abstract}
\end{titlepage}
\setcounter{page}{2}
%

High-energy heavy-ion collisions offer the unique possibility to study nuclear matter under extreme conditions, in particular the deconfined phase (Quark--Gluon Plasma, QGP~\cite{Cabibbo:1975ig,Shuryak:1978ij,McLerran:1980pk}), which has been predicted by lattice QCD~\cite{Laermann:2003cv}. The transverse momentum (\pt) distributions and yields of identified particles are instrumental to the study of the collective and thermal properties of this matter. 
Results from lower energies, in particular from the Relativistic Heavy-Ion Collider (RHIC, \snn~= 200~GeV), have shown that the bulk matter created in high-energy nuclear reactions can be quantitatively described in terms of hydrodynamic models. The initial hot and dense partonic matter rapidly expands and cools down, ultimately undergoing a transition to a hadron gas phase~\cite{Muller:2006ee}.  The observed particle abundances were described in terms of thermal models. Relative particle abundances in thermal and chemical equilibrium are governed mainly by two parameters, the chemical freeze-out temperature \Tch~\ and the baryochemical potential \muB, where the latter describes the net baryon content of the system~\cite{Andronic:2011yq,Andronic:2008gu,Becattini:2010lb,Cleymans:1998fq}. Measured particle yields in heavy-ion collisions at RHIC, as well as SPS and AGS, are consistent with equilibrium populations, allowing the extraction of both model parameters from fits to the measured particle ratios~\cite{Andronic:2011yq,Andronic:2008gu,Cleymans:1998fq,Andronic:2005yp}.  It has been argued that interactions modifying the relative abundances of particle species are negligible in the hadronic phase~\cite{Rapp:2000gy,Greiner:2000tu}, and that \Tch\ can be linked to the phase transition temperature~\cite{BraunMunzinger:2003zz}.
Particle momentum distributions reflect the conditions later in the evolution, at the ``kinetic freeze-out'' from the hadron gas phase, when elastic interactions end~\cite{Huovinen:2006jp}.
The \pt\ distributions encode information about the collective transverse expansion (radial flow) and the temperature \Tfo\ at the kinetic freeze-out~\cite{Schnedermann:1993ws,Heinz:2004qz}. The collective expansion is driven by internal pressure gradients and quantified by the average transverse velocity \avbT.
Based on the success of the thermal and hydrodynamic models and based on the trend of the model parameters with \snn, predictions were formulated for higher energy~\cite{Andronic:2008gu,Cleymans:2006xj}. With the advent of the LHC, it became important to check these expectations in the new energy regime. 
In this Letter, we present the first results on $\pi$, K, and p production in 0-5\% central \PbPb\ collisions at \snn~=~2.76 TeV, measured by the ALICE
       experiment at the LHC. Previous results in pp collisions are reported in~\cite{Aamodt:2011zj}. The measurement spans the \pt\ range from $\sim$0.1 up to $\sim$4.5~GeV/$c$ at central rapidity ($\left|y\right|<0.5$). 

 The central tracking and particle identification (PID) detectors cover the pseudorapidity window $\left| \eta \right| < 0.9$\warn{o 0.8?} and include, from the innermost outwards, the Inner Tracking System (ITS), the Time Projection Chamber (TPC), the Transition Radiation Detector (TRD) and the Time-Of-Flight array (TOF)~\cite{Alessandro:2006yt,Aamodt:2008zz}. The central detectors are operated in a 0.5~T solenoidal field. The moderate field, together with a low material budget ($X/X_0 \approx 0.1$ for a track going through the TPC) permits the reconstruction of low \pt\ tracks.  The data sample consists of 4 M minimum bias events. 
A pair of forward scintillator hodoscopes, the \VZERO\ detectors ($2.8 < \eta < 5.1$ and $-3.7 < \eta < -1.7$), was used for triggering and for centrality determination~\cite{Aamodt:2010cz,Aamodt:2010pb,centrality-paper}.  The data were collected using a minimum bias trigger requiring a combination of hits in the ITS and in the \VZERO, a condition fully efficient for the event sample discussed here~\cite{centrality-paper}.  The 0-5\% most central collisions were selected using the signal amplitudes measured in the \VZERO, fitted with a Glauber model~\cite{centrality-paper,Toia:2011rta}.  Background events caused by beam-gas interactions, parasitic collisions and electromagnetic processes were rejected using timing cuts on the \VZERO\ and two neutron Zero-Degree Calorimeters located on either side of the interaction point, at 114 m distance~\cite{Aamodt:2010cz,Aamodt:2010pb,centrality-paper}.
The measurement was performed in three independent analyses, each one focusing on a sub-range of the total \pt~distribution,  exploiting the capabilities of the individual detectors and specific techniques to optimize the signal extraction. 
The ITS is a six-layered silicon detector. The two innermost layers, based on silicon pixels, are also used in the online trigger as mentioned above. 
The four outer layers, consisting of drift and strip detectors, provide identification via the specific energy loss.
Using the ITS as a standalone tracker enables
reconstruction and identification of low-momentum particles that do not reach the TPC, in the \pt\   ranges 0.1-0.6, 0.2-0.5, 0.3-0.6~GeV/$c$ for $\pi$, K, and p. 
For each track, at least three \dedx\ samples were required. Only the lowest two were used in a truncated mean procedure, leading to an $\sim$10\% resolution. The particle identity was assigned according to the distance from the expected energy loss curves, weighted by the resolution.  This procedure results in asymmetric ranges around the curves for $\pi$, K, and p, reflecting the Landau tails in
the detector response, which are not fully suppressed by the truncated mean.  An additional 2~$\sigma$ cut was applied in the case of pions, to remove the contamination from electrons. The residual misidentification ($< 10\%$ for kaons and negligible for pions and protons) is corrected using Monte Carlo. 
The other analyses combined the tracking information from the ITS, TPC and TRD (``global tracks'')~\cite{Alessandro:2006yt}. 
The TPC identifies particles via the specific energy loss in the Ne-CO$_2$-N$_2$ (85:10:5) gas mixture. Up to 159 samples are measured, but only the  lowest 60\% are used in the analysis. This truncated mean procedure leads to a Gaussian distribution with an $\sim$6.5\% resolution. The TOF is placed at a radius of 370--399~cm. It measures the time-of-flight of the particles, allowing hadron identification at higher \pt. With a total time resolution of about 85~ps, PID is possible up to \pt~=~3~GeV/$c$ for $\pi$ and K, and 4.5~GeV/$c$ for p.
In the intermediate \pt\ range (0.2-1.2~GeV/$c$, 0.25-1.2~GeV/$c$, 0.45-1.8~GeV/$c$, for $\pi$ and K, and p), the identification was based on the combined TPC and TOF signals. The ranges were chosen such that the contamination from misidentification of other species is negligible. 
It was required that the particles are within 3~$\sigma$ from the expected \dedx\ and time-of-flight values. The TOF information was included starting at \pt~=~0.65, 0.6, 0.8~GeV/$c$ for $\pi$, K, and p.  In the third analysis, for $\pt > 0.5$~GeV/$c$, the TOF signal alone was used for identification. The time-of-flight distribution was fitted for each \pt\ bin with data-derived templates for $\pi$, K, and p, allowing to extract the particle yields when the separation is as low as $\sim$2~$\sigma$~\cite{prc-spectra}. This fit was repeated for each mass hypothesis, after applying the selection $\left|y\right| < 0.5$ based on the mass assumption under study~\cite{prc-spectra}. 

In this Letter, results for ``primary'' particles are presented, defined as prompt particles produced in the collision, including decay products, except those from weak decays of strange particles.
Both ITS standalone and global tracks provide very good transverse impact parameter resolution relative to the primary vertex (\dcaxy), of order 200~$\mu$m at \pt~=~300~MeV/$c$ and 35~$\mu$m at \pt~=~5~GeV/$c$, allowing to separate primary and secondary particles.
 The residual contamination was estimated from data by fitting the \dcaxy\ distribution with three Monte Carlo templates: primary particles, secondaries from material and secondaries from weak decays~\cite{Kalweit:2011if,Aamodt:2011zj}. This contamination can reach 35\% at \pt~=~300 MeV/$c$ for protons. It quickly decreases with increasing \pt, becoming negligible at \pt~$\sim$2.5~GeV/$c$. 
The efficiency correction and the templates used in the correction procedure were computed with 1~M Monte Carlo events, using the Hijing~\cite{hijing} event generator (tuned to reproduce the \dNdeta\ measured for central collisions~\cite{Aamodt:2010pb}), transported through a GEANT3~\cite{Geant:1994zzo} model of the detector.
The results of the three analyses were consistent in the regions of overlap and therefore combined using the (largely independent) systematic uncertainties as weights.

\begin{table}
\centering \caption{Main sources of systematic uncertainty. See text for details.} \label{tab:syst}
\begin{tabular}{l|cc|cc|cc}
\hline
\hline
effect & \multicolumn{2}{c|}{$\pi^{\pm}$} & \multicolumn{2}{c|}{K$^{\pm}$} & \multicolumn{2}{c}{p and \pbar} \\ [3pt]
\hline
\pt\ range (GeV/$c$) & 0.1 & 3 & 0.2 & 3 & 0.35 & 4.5\\
\hline

correction for &  \multirow{2}{*}{1.5\%} & \multirow{2}{*}{1\%} & \multicolumn{2}{c|}{\multirow{2}{*}{negl.}} & \multirow{2}{*}{4\%} & \multirow{2}{*}{1\%} \\ 
secondaries & & & & & & \\[3pt]

material & \multirow{2}{*}{5\%} & \multirow{2}{*}{negl.} & \multirow{2}{*}{3\%} & \multirow{2}{*}{negl.} & \multirow{2}{*}{3\%} & \multirow{2}{*}{negl.} \\
budget & & & & & & \\ [3pt]

hadronic      &  \multirow{2}{*}{2\%} & \multirow{2}{*}{1\%} &   \multirow{2}{*}{4\%} & \multirow{2}{*}{1\%} &  
6\% & 1\% (\pbar) \\
interactions & & & & &  4\% & negl. (p) \\  [3pt]  

\hline
\pt\ range (GeV/$c$) & 0.1 & 0.5 & 0.2 & 0.5 & 0.35 & 0.65\\
\hline

ITS tracking &  \multicolumn{2}{c|}{\multirow{2}{*}{10\%}} & \multicolumn{2}{c|}{\multirow{2}{*}{10\%}} & \multicolumn{2}{c}{\multirow{2}{*}{10\%}} \\
efficiency &&&&&& \\ [3pt]

ITS PID &  \multicolumn{2}{c|}{2\%} & \multicolumn{2}{c|}{4\%} & \multicolumn{2}{c}{4.5\%} \\ [3pt]

\hline
\pt\ range (GeV/$c$) & 0.3 & 0.65 & 0.3 & 0.6 & 0.5 & 0.8\\
\hline

global tracking&  \multicolumn{2}{c|}{\multirow{2}{*}{4\%}} & \multicolumn{2}{c|}{\multirow{2}{*}{4\%}} & \multicolumn{2}{c}{\multirow{2}{*}{4\%}} \\
efficiency &&&&&& \\ [3pt]

TPC PID & \multicolumn{2}{c|}{3\%} & \multicolumn{2}{c|}{5\%} & \multicolumn{2}{c}{1.5\%} \\ [3pt]

\hline
\pt\ range (GeV/$c$) & 0.5 & 3 & 0.5 & 3 & 0.5 & 4.5\\
\hline

TOF matching  & \multicolumn{2}{c|}{\multirow{2}{*}{3\%}} & \multicolumn{2}{c|}{\multirow{2}{*}{6\%}} & \multicolumn{2}{c}{\multirow{2}{*}{3\%}} \\ 
efficiency & & & & & & \\ [3pt]
TOF PID & 2\% & 7\% & 3\% & 15\% & 5\% & 25\% \\ [3pt]
\hline\noalign{\smallskip}
\hline\noalign{\smallskip}

\end{tabular}
\end{table}

The main sources of systematic uncertainties are summarized in Table~\ref{tab:syst}.
The uncertainties due to the secondary subtraction procedure were estimated for all analyses varying the range of the \dcaxy\ fit, using different track selections (for instance using TPC-only tracks), applying different cuts on the longitudinal \dcaz\ and varying the composition of the Monte Carlo templates used in the fit.
The uncertainty on the energy loss correction was estimated in a simulation with the material budget scaled by $\pm$7\%. In order to account for the uncertainties due to poorly-known hadronic interactions with the detector material, different transport codes (GEANT3, GEANT4~\cite{Agostinelli:2002hh} and FLUKA~\cite{Battistoni:2007zzb}) were tested. The interaction cross section used in the models for $\pi$, K, and p were separately validated by comparison with the few existing measurements~\cite{Lee:2002eq,Friedman:1997eq,Ashery:1981tq,Carlson199693,Bendiscioli:1994uv}. 
The main systematic uncertainty on the ITS standalone analysis comes from the tracking, due to the high occupancy and small number of tracking points. This was estimated from data using global tracks as a reference. 
The other systematic uncertainties are smaller, and include the effect of the magnetic field configuration (\EcrossB\ effect), of the track selection and of the PID cuts.
Similarly, the uncertainties related to the tracking efficiency in the TPC were investigated comparing different sets of tracks in data and Monte Carlo and by a variation of the track cuts. 
The uncertainty related to the combined TPC/TOF PID procedure was estimated varying the PID cut between 2 and 5~$\sigma$. 
The tracks reaching the TOF detector have to cross a substantial amount of additional material ($X/X_0 \approx 0.23$), mostly due to the TRD~\cite{Aamodt:2008zz}. Since the TRD was not fully installed in 2010, the analysis was repeated for regions in azimuth with and without installed TRD modules, allowing one to determine the uncertainty due to the aditional material. 
The systematics related to the PID extraction in the TOF analysis were estimated varying the parameters of the expected sources by $\pm$10\%.

\begin{figure}[tb]
  \centering
  \includegraphics[width=0.48\textwidth]{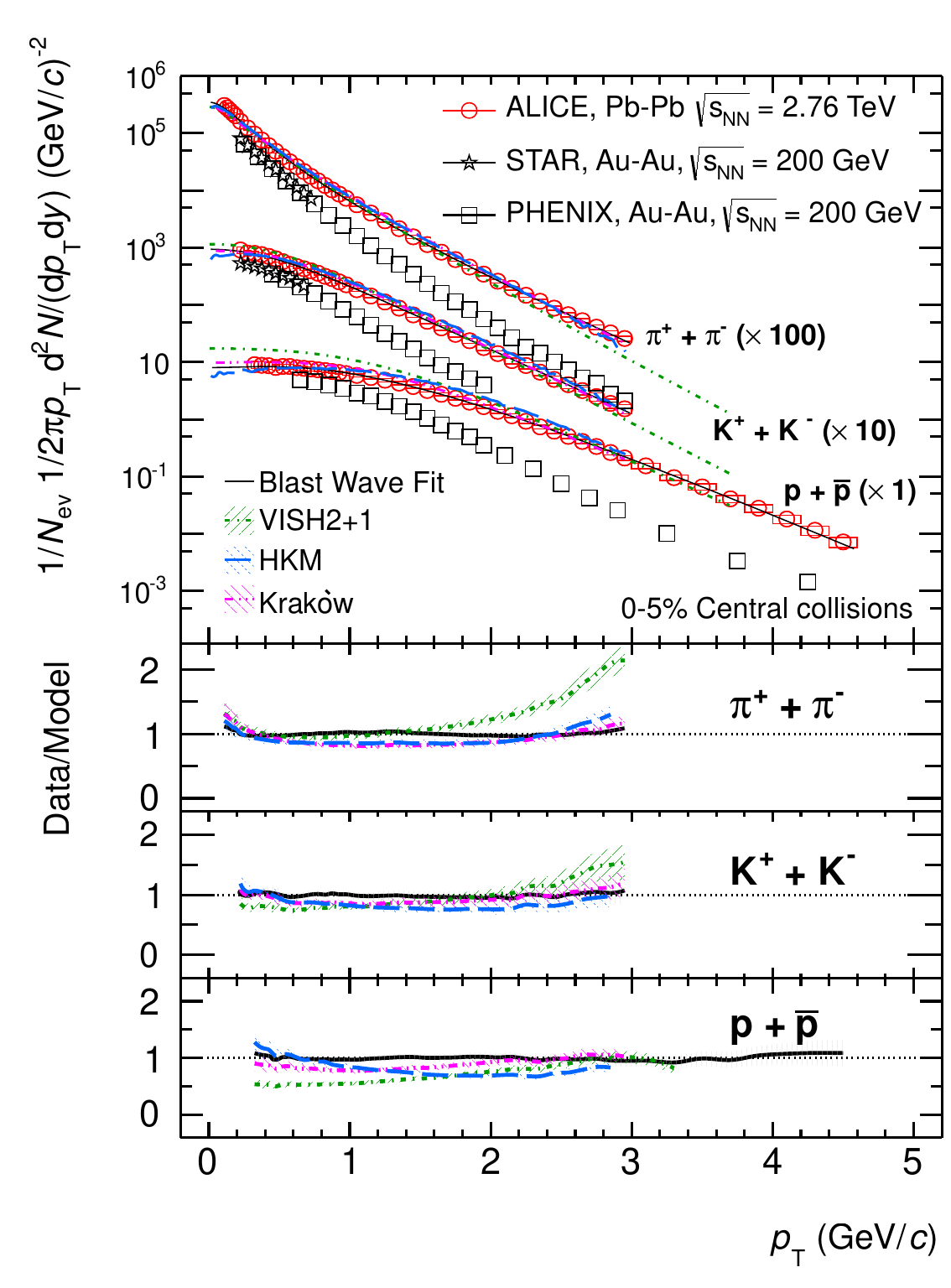}
\caption{(color online) Transverse momentum distributions  of the sum of positive and negative particles (box: systematic errors; statistical errors smaller than the symbol for most data points), fitted individually with a blast wave function, compared to RHIC data and hydrodynamic models.}
  \label{fig:spectra-central}
\end{figure}

The \pt\ distributions of positive and negative particles were found to be compatible within systematic errors, we therefore show results for summed charge states in Fig.~\ref{fig:spectra-central}. 
The spectra are compared to RHIC results in \AuAu\ collisions at \snn~=~200~GeV~\cite{Abelev:2008ez,Adler:2003cb} and to hydrodynamic models. The  spectral shapes show a significant change from RHIC to LHC energies, having a distinctly harder distribution. 
Within hydrodynamic models, this indicates a significantly stronger radial flow.
In the range $\pt < 1.5$~GeV/$c$ VISH2+1~\cite{Shen:2011eg}, a viscous hydrodynamic model, reproduces fairly well the pion and kaon distributions, but misses the protons, both in shape and absolute abundance. In this model, the particle yields are taken to be thermal at $\Tch = 165$~MeV (see below). The difference is possibly due to the lack of an explicit description of the hadronic phase in the model. This interpretation is supported by the comparison with HKM~\cite{Karpenko:2011qn,Karpenko:2012yf}, a similar model in which, after the hydrodynamic phase, particles are injected into a hadronic cascade model (UrQMD~\cite{Bass:1998ca,Bleicher:1999xi}), which further transports them until final decoupling. The hadronic phase builds additional radial flow, mostly due to elastic interactions, and affects particle ratios due to inelastic interactions. HKM yields a better description of the data. At the LHC, hadronic final state interactions, and  in particular antibaryon-baryon annihilation, may therefore be an important ingredient for the description of particle yields~\cite{Steinheimer:2012rd,Karpenko:2012yf}, contradicting the scenario of negligible abundance-changing processes in the hadronic phase. 
The third model shown in Fig.~\ref{fig:spectra-central} (Krak\'ow~\cite{Bozek:2011ua,Bozek:2011gq}) 
introduces non-equilibrium corrections due to viscosity at the transition from the hydrodynamic description to particles, which change the effective \Tch, leading to a good agreement with the data. In the region $\pt \lesssim 3$~GeV/$c$ (Krak\'ow) and $\pt \lesssim 1.5$~GeV/$c$ (HKM) the last two models reproduce the experimental data within $\sim$20\%, supporting a hydrodynamic interpretation of the transverse momentum spectra at the LHC. These models also describe correctly other features of the space-time evolution of the system, as measured by ALICE with charged pion correlations~\cite{Aamodt:2011mr}. 
In order to quantify the kinetic freeze-out parameters at \snn~=~2.76~TeV, we performed a  combined fit with a blast wave function~\cite{Schnedermann:1993ws}, with the freeze out temperature \Tfo, the average transverse velocity \avbT\ and the exponent of the velocity profile as free parameters. It should be noted that the value of the \Tfo\ parameter extracted from the fit is sensitive to the fit range used for the pions, because of the large  contribution from resonance decays (mostly at low \pt), which tend to reduce \Tfo. For this reason, the  \pt\ ranges 0.5-1 GeV/$c$, 0.2-1.5 GeV/$c$, 0.3-3 GeV/$c$ for $\pi$, K, and p were used. 
These hydro-motivated fits do not replace a full hydrodynamic calculation, but allow one to compare with a few parameters the measurements of different experiments. The data are well described by the combined blast wave fit with a collective radial flow velocity 
$\avbT = 0.65 \pm 0.02$, and a kinetic freeze-out temperature of $\Tfo = 95 \pm 10$~MeV. 
 As compared to fits to central \AuAu\ collisions at \snn~=~200~GeV/$c$, in similar \pt\ ranges~\cite{Abelev:2008ez,Adams:2005dq}, \avbT\ at the LHC is $\sim$10\% higher while  \Tfo\  is comparable within errors.

\begin{figure}[tb]
  \centering
 \includegraphics[width=0.49\textwidth]{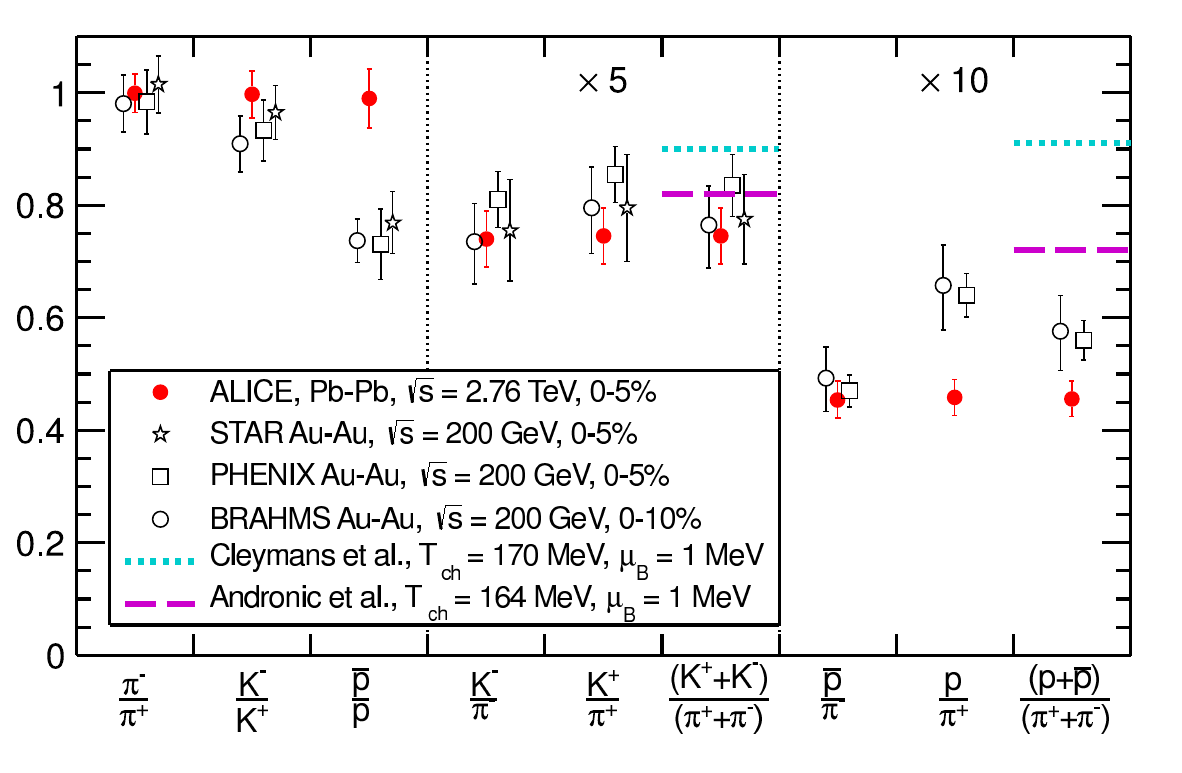}
\caption{(color online) Mid-rapidity particle ratios, compared to RHIC results~\cite{Abelev:2008ez,Adler:2003cb,Arsene:2005mr} and predictions from thermal models~\cite{Andronic:2008gu,Cleymans:2006xj} for central Pb-Pb collisions at the LHC (combined statistical and systematic errors).}
  \label{fig:ratios}
\end{figure}

The mid-rapidity ($\left|y\right|<0.5$) \pt-integrated particle yields were extracted by fitting the $\pi$, K, and p spectra individually with a blast wave function, in order to extrapolate to zero \pt. The individual fits are shown in Fig.~\ref{fig:spectra-central} as solid curves; the fraction of extrapolated yield is small: about 7\%, 6\%, and 4\% for $\pi$, K, and p. Its uncertainty was estimated using different fit functions~\cite{prc-spectra}. 
The particle ratios are compared in Fig.~\ref{fig:ratios} to results at \snn~=~200~GeV~\cite{Abelev:2008ez,Adler:2003cb,Arsene:2005mr}  and to the predictions from thermal models, using $\muB = 1~\mathrm{MeV}$ and a \Tch\ of 164~MeV~\cite{Andronic:2008gu} or 170~MeV~\cite{Cleymans:2006xj}. The value for $\muB$ is based on extrapolation from lower energy data. $\Tch$ was found to be constant above a center-of-mass energy of a few ten GeV, so the value obtained from fits to RHIC data was used. 
The systematic uncertainties on the particle ratios were computed taking into account the correlated sources of uncertainty (mainly due to PID and extrapolation for different particles and to the tracking efficiency for anti-particle over particle ratios). In the following we quote the total error for the ratios, as the statistical error is negligible. 
The anti-particle/particle ratios are all unity within errors, consistent with a vanishing baryochemical potential \muB. 
In order to minimize the sensitivity to \muB, the ratios $\kpi=(\mathrm{K}^+ + \mathrm{K}^-)/(\pi^+ + \pi^-)$\ and $\ppi=(\mathrm{p} + \mathrm{\bar p})/(\pi^+ + \pi^-)$ are also shown. The ratio \kpi~=~$0.149 \pm 0.010$, is similar to the lower energy values and agrees with the expectations from the thermal model~\cite{Andronic:2008gu}.
However, the ratio \ppi$~= 0.046 \pm 0.003$, is significantly lower than expected, by a factor $\sim$1.5--1.9 ($\ppi \simeq 0.07-0.09$ for \cite{Andronic:2008gu} and \cite{Cleymans:2006xj} respectively). The two models differ mainly in the hadron mass spectrum implementation, but were both successful in describing RHIC data. 
The comparison with RHIC data also hints at a slight decrease of the \ppi\ ratio with energy (by a factor $\sim$1.2), while essentially no change was predicted.
The thermal models proved to be very successful over a wide range of energies (from \snn=~2~GeV to \snn~=~200~GeV~\cite{Cleymans:1998fq,Andronic:2008gu,Andronic:2011yq,Andronic:2005yp}): such a large difference  for one of the most abundantly produced particle species was therefore unexpected. 
In retrospect, some disagreement between data and the thermal model is also apparent in the RHIC data,
with the proton measurements being about 20\% lower than predictions~\cite{Andronic:2011yq,Adams:2005dq,Aggarwal:2010ig}. 
However, this difference was not considered to be significant, 
because of the differences between model implementations, model uncertainties~\cite{Becattini:2010sk}
and experimental uncertainties in the subtraction of secondary particles in the RHIC experiments. This issue will likely be clarified by a thermal analysis including strange and multi-strange baryons at the LHC.
Current speculations are that final state interactions in the hadronic phase, in particular via the large cross section channel for antibaryon-baryon annihilation~\cite{Steinheimer:2012rd}, could explain the significant deviation from the usual thermal ratios. A similar conclusion is implied by the HKM model, where $\ppi = 0.052$, consistent with our measurement~\cite{Karpenko:2012yf}. 
An alternative scenario conjectures the existence of flavor and mass dependent pre-hadronic bound states in the QGP phase, as suggested by recent lattice QCD calculation and QCD-inspired models~\cite{Ratti:2011au,Bellwied:2012kh}.

In summary, we presented the first measurements of $\pi$, K, and p production in central Pb-Pb collisions at \snn~=~2.76~TeV at the LHC. 
The \pt\ distributions are harder than previously measured at RHIC. They are
well described by hydrodynamic models including a refined description of the late fireball stages.
Fitting the spectra with a hydro-inspired blast wave model results in the highest radial flow parameter ever measured, $\avbT = 0.65 \pm 0.02$.  The integrated particle ratios were compared with expectations from thermal models. While the \kpi\ ratio was found to agree with these expectations, \ppi\ is a factor $\gtrsim 1.5$ lower. 

\newenvironment{acknowledgement}{\relax}{\relax}
\begin{acknowledgement}
\section*{Acknowledgments}
We are grateful to P. Bozek, U. Heinz, Y. Karpenko, C. Shen, Y. Sinyukov, H. Song and  K. Werner for providing the theoretical calculations and for the useful discussion and to colleagues from the BRAHMS, PHENIX and STAR collaborations for the helpful discussions and clarifications on their measurements.


The ALICE collaboration would like to thank all its engineers and technicians for their invaluable contributions to the construction of the experiment and the CERN accelerator teams for the outstanding performance of the LHC complex.
The ALICE collaboration acknowledges the following funding agencies for their support in building and
running the ALICE detector:
Calouste Gulbenkian Foundation from Lisbon and Swiss Fonds Kidagan, Armenia;
Conselho Nacional de Desenvolvimento Cient\'{\i}fico e Tecnol\'{o}gico (CNPq), Financiadora de Estudos e Projetos (FINEP),
Funda\c{c}\~{a}o de Amparo \`{a} Pesquisa do Estado de S\~{a}o Paulo (FAPESP);
National Natural Science Foundation of China (NSFC), the Chinese Ministry of Education (CMOE)
and the Ministry of Science and Technology of China (MSTC);
Ministry of Education and Youth of the Czech Republic;
Danish Natural Science Research Council, the Carlsberg Foundation and the Danish National Research Foundation;
The European Research Council under the European Community's Seventh Framework Programme;
Helsinki Institute of Physics and the Academy of Finland;
French CNRS-IN2P3, the `Region Pays de Loire', `Region Alsace', `Region Auvergne' and CEA, France;
German BMBF and the Helmholtz Association;
Hungarian OTKA and National Office for Research and Technology (NKTH);
Department of Atomic Energy and Department of Science and Technology of the Government of India;
Istituto Nazionale di Fisica Nucleare (INFN) of Italy;
MEXT Grant-in-Aid for Specially Promoted Research, Ja\-pan;
Joint Institute for Nuclear Research, Dubna;
 %
National Research Foundation of Korea (NRF);
CONACYT, DGAPA, M\'{e}xico, ALFA-EC and the HELEN Program (High-Energy physics Latin-American--European Network);
Stichting voor Fundamenteel Onderzoek der Materie (FOM) and the Nederlandse Organisatie voor Wetenschappelijk Onderzoek (NWO), Netherlands;
Research Council of Norway (NFR);
Polish Ministry of Science and Higher Education;
National Authority for Scientific Research - NASR (Autoritatea Na\c{t}ional\u{a} pentru Cercetare \c{S}tiin\c{t}ific\u{a} - ANCS);
Federal Agency of Science of the Ministry of Education and Science of Russian Federation, International Science and
Technology Center, Russian Academy of Sciences, Russian Federal Agency of Atomic Energy, Russian Federal Agency for Science and Innovations and CERN-INTAS;
Ministry of Education of Slovakia;
CIEMAT, EELA, Ministerio de Educaci\'{o}n y Ciencia of Spain, Xunta de Galicia (Conseller\'{\i}a de Educaci\'{o}n),
CEA\-DEN, Cubaenerg\'{\i}a, Cuba, and IAEA (International Atomic Energy Agency);
The Ministry of Science and Technology and the National Research Foundation (NRF), South Africa;
Swedish Reseach Council (VR) and Knut $\&$ Alice Wallenberg Foundation (KAW);
Ukraine Ministry of Education and Science;
United Kingdom Science and Technology Facilities Council (STFC);
The United States Department of Energy, the United States National
Science Foundation, the State of Texas, and the State of Ohio.
\end{acknowledgement}


%
%
\bibliographystyle{h-physrev}
\bibliography{prlSpectra2}
\newpage
%
%
\appendix
\section{The ALICE Collaboration}
\label{app:collab}

\begingroup
\small
\begin{flushleft}
B.~Abelev\Irefn{org1234}\And
J.~Adam\Irefn{org1274}\And
D.~Adamov\'{a}\Irefn{org1283}\And
A.M.~Adare\Irefn{org1260}\And
M.M.~Aggarwal\Irefn{org1157}\And
G.~Aglieri~Rinella\Irefn{org1192}\And
A.G.~Agocs\Irefn{org1143}\And
A.~Agostinelli\Irefn{org1132}\And
S.~Aguilar~Salazar\Irefn{org1247}\And
Z.~Ahammed\Irefn{org1225}
% ...
\renewcommand\labelenumi{\textsuperscript{\theenumi}~}
\section*{Affiliation notes}
\renewcommand\theenumi{\roman{enumi}}
\begin{Authlist}
\item \Adef{M.V.Lomonosov Moscow State University, D.V.Skobeltsyn Institute of Nuclear Physics, Moscow, Russia}Also at: M.V.Lomonosov Moscow State University, D.V.Skobeltsyn Institute of Nuclear Physics, Moscow, Russia
\end{Authlist}
\section*{Collaboration Institutes}
\renewcommand\theenumi{\arabic{enumi}~}
\begin{Authlist}
\item \Idef{org1279}Benem\'{e}rita Universidad Aut\'{o}noma de Puebla, Puebla, Mexico
\item \Idef{org1220}Bogolyubov Institute for Theoretical Physics, Kiev, Ukraine
\item \Idef{org1262}Budker Institute for Nuclear Physics, Novosibirsk, Russia
\item \Idef{org1292}California Polytechnic State University, San Luis Obispo, California, United States
\item \Idef{org1329}Central China Normal University, Wuhan, China
\item \Idef{org14939}Centre de Calcul de l'IN2P3, Villeurbanne, France
\item \Idef{org1197}Centro de Aplicaciones Tecnol\'{o}gicas y Desarrollo Nuclear (CEADEN), Havana, Cuba
\item \Idef{org1242}Centro de Investigaciones Energ\'{e}ticas Medioambientales y Tecnol\'{o}gicas (CIEMAT), Madrid, Spain
\item \Idef{org1244}Centro de Investigaci\'{o}n y de Estudios Avanzados (CINVESTAV), Mexico City and M\'{e}rida, Mexico
\item \Idef{org1335}Centro Fermi -- Centro Studi e Ricerche e Museo Storico della Fisica ``Enrico Fermi'', Rome, Italy
\item \Idef{org17347}Chicago State University, Chicago, United States
\item \Idef{org1288}Commissariat \`{a} l'Energie Atomique, IRFU, Saclay, France
\item \Idef{org1294}Departamento de F\'{\i}sica de Part\'{\i}culas and IGFAE, Universidad de Santiago de Compostela, Santiago de Compostela, Spain
\item \Idef{org1106}Department of Physics Aligarh Muslim University, Aligarh, India
\item \Idef{org1121}Department of Physics and Technology, University of Bergen, Bergen, Norway
\item \Idef{org1162}Department of Physics, Ohio State University, Columbus, Ohio, United States
\item \Idef{org1300}Department of Physics, Sejong University, Seoul, South Korea
\item \Idef{org1268}Department of Physics, University of Oslo, Oslo, Norway
\item \Idef{org1132}Dipartimento di Fisica dell'Universit\`{a} and Sezione INFN, Bologna, Italy
\item \Idef{org1270}Dipartimento di Fisica dell'Universit\`{a} and Sezione INFN, Padova, Italy
\item \Idef{org1315}Dipartimento di Fisica dell'Universit\`{a} and Sezione INFN, Trieste, Italy
\item \Idef{org1145}Dipartimento di Fisica dell'Universit\`{a} and Sezione INFN, Cagliari, Italy
\item \Idef{org1312}Dipartimento di Fisica dell'Universit\`{a} and Sezione INFN, Turin, Italy
\item \Idef{org1285}Dipartimento di Fisica dell'Universit\`{a} `La Sapienza' and Sezione INFN, Rome, Italy
\item \Idef{org1154}Dipartimento di Fisica e Astronomia dell'Universit\`{a} and Sezione INFN, Catania, Italy
\item \Idef{org1290}Dipartimento di Fisica `E.R.~Caianiello' dell'Universit\`{a} and Gruppo Collegato INFN, Salerno, Italy
\item \Idef{org1103}Dipartimento di Scienze e Innovazione Tecnologica dell'Universit\`{a} del Piemonte Orientale and Gruppo Collegato INFN, Alessandria, Italy
\item \Idef{org1114}Dipartimento Interateneo di Fisica `M.~Merlin' and Sezione INFN, Bari, Italy
\item \Idef{org1237}Division of Experimental High Energy Physics, University of Lund, Lund, Sweden
\item \Idef{org1192}European Organization for Nuclear Research (CERN), Geneva, Switzerland
\item \Idef{org1227}Fachhochschule K\"{o}ln, K\"{o}ln, Germany
\item \Idef{org1122}Faculty of Engineering, Bergen University College, Bergen, Norway
\item \Idef{org1136}Faculty of Mathematics, Physics and Informatics, Comenius University, Bratislava, Slovakia
\item \Idef{org1274}Faculty of Nuclear Sciences and Physical Engineering, Czech Technical University in Prague, Prague, Czech Republic
\item \Idef{org1229}Faculty of Science, P.J.~\v{S}af\'{a}rik University, Ko\v{s}ice, Slovakia
\item \Idef{org1184}Frankfurt Institute for Advanced Studies, Johann Wolfgang Goethe-Universit\"{a}t Frankfurt, Frankfurt, Germany
\item \Idef{org1215}Gangneung-Wonju National University, Gangneung, South Korea
\item \Idef{org1212}Helsinki Institute of Physics (HIP) and University of Jyv\"{a}skyl\"{a}, Jyv\"{a}skyl\"{a}, Finland
\item \Idef{org1203}Hiroshima University, Hiroshima, Japan
\item \Idef{org1254}Indian Institute of Technology Bombay (IIT), Mumbai, India
\item \Idef{org36378}Indian Institute of Technology Indore (IIT), Indore, India
\item \Idef{org1266}Institut de Physique Nucl\'{e}aire d'Orsay (IPNO), Universit\'{e} Paris-Sud, CNRS-IN2P3, Orsay, France
\item \Idef{org1277}Institute for High Energy Physics, Protvino, Russia
\item \Idef{org1249}Institute for Nuclear Research, Academy of Sciences, Moscow, Russia
\item \Idef{org1320}Nikhef, National Institute for Subatomic Physics and Institute for Subatomic Physics of Utrecht University, Utrecht, Netherlands
\item \Idef{org1250}Institute for Theoretical and Experimental Physics, Moscow, Russia
\item \Idef{org1230}Institute of Experimental Physics, Slovak Academy of Sciences, Ko\v{s}ice, Slovakia
\item \Idef{org1127}Institute of Physics, Bhubaneswar, India
\item \Idef{org1275}Institute of Physics, Academy of Sciences of the Czech Republic, Prague, Czech Republic
\item \Idef{org1139}Institute of Space Sciences (ISS), Bucharest, Romania
\item \Idef{org27399}Institut f\"{u}r Informatik, Johann Wolfgang Goethe-Universit\"{a}t Frankfurt, Frankfurt, Germany
\item \Idef{org1185}Institut f\"{u}r Kernphysik, Johann Wolfgang Goethe-Universit\"{a}t Frankfurt, Frankfurt, Germany
\item \Idef{org1177}Institut f\"{u}r Kernphysik, Technische Universit\"{a}t Darmstadt, Darmstadt, Germany
\item \Idef{org1256}Institut f\"{u}r Kernphysik, Westf\"{a}lische Wilhelms-Universit\"{a}t M\"{u}nster, M\"{u}nster, Germany
\item \Idef{org1246}Instituto de Ciencias Nucleares, Universidad Nacional Aut\'{o}noma de M\'{e}xico, Mexico City, Mexico
\item \Idef{org1247}Instituto de F\'{\i}sica, Universidad Nacional Aut\'{o}noma de M\'{e}xico, Mexico City, Mexico
\item \Idef{org23333}Institut of Theoretical Physics, University of Wroclaw
\item \Idef{org1308}Institut Pluridisciplinaire Hubert Curien (IPHC), Universit\'{e} de Strasbourg, CNRS-IN2P3, Strasbourg, France
\item \Idef{org1182}Joint Institute for Nuclear Research (JINR), Dubna, Russia
\item \Idef{org1143}KFKI Research Institute for Particle and Nuclear Physics, Hungarian Academy of Sciences, Budapest, Hungary
\item \Idef{org1199}Kirchhoff-Institut f\"{u}r Physik, Ruprecht-Karls-Universit\"{a}t Heidelberg, Heidelberg, Germany
\item \Idef{org20954}Korea Institute of Science and Technology Information, Daejeon, South Korea
\item \Idef{org1160}Laboratoire de Physique Corpusculaire (LPC), Clermont Universit\'{e}, Universit\'{e} Blaise Pascal, CNRS--IN2P3, Clermont-Ferrand, France
\item \Idef{org1194}Laboratoire de Physique Subatomique et de Cosmologie (LPSC), Universit\'{e} Joseph Fourier, CNRS-IN2P3, Institut Polytechnique de Grenoble, Grenoble, France
\item \Idef{org1187}Laboratori Nazionali di Frascati, INFN, Frascati, Italy
\item \Idef{org1232}Laboratori Nazionali di Legnaro, INFN, Legnaro, Italy
\item \Idef{org1125}Lawrence Berkeley National Laboratory, Berkeley, California, United States
\item \Idef{org1234}Lawrence Livermore National Laboratory, Livermore, California, United States
\item \Idef{org1251}Moscow Engineering Physics Institute, Moscow, Russia
\item \Idef{org1140}National Institute for Physics and Nuclear Engineering, Bucharest, Romania
\item \Idef{org1165}Niels Bohr Institute, University of Copenhagen, Copenhagen, Denmark
\item \Idef{org1109}Nikhef, National Institute for Subatomic Physics, Amsterdam, Netherlands
\item \Idef{org1283}Nuclear Physics Institute, Academy of Sciences of the Czech Republic, \v{R}e\v{z} u Prahy, Czech Republic
\item \Idef{org1264}Oak Ridge National Laboratory, Oak Ridge, Tennessee, United States
\item \Idef{org1189}Petersburg Nuclear Physics Institute, Gatchina, Russia
\item \Idef{org1170}Physics Department, Creighton University, Omaha, Nebraska, United States
\item \Idef{org1157}Physics Department, Panjab University, Chandigarh, India
\item \Idef{org1112}Physics Department, University of Athens, Athens, Greece
\item \Idef{org1152}Physics Department, University of Cape Town, iThemba LABS, Cape Town, South Africa
\item \Idef{org1209}Physics Department, University of Jammu, Jammu, India
\item \Idef{org1207}Physics Department, University of Rajasthan, Jaipur, India
\item \Idef{org1200}Physikalisches Institut, Ruprecht-Karls-Universit\"{a}t Heidelberg, Heidelberg, Germany
\item \Idef{org1325}Purdue University, West Lafayette, Indiana, United States
\item \Idef{org1281}Pusan National University, Pusan, South Korea
\item \Idef{org1176}Research Division and ExtreMe Matter Institute EMMI, GSI Helmholtzzentrum f\"ur Schwerionenforschung, Darmstadt, Germany
\item \Idef{org1334}Rudjer Bo\v{s}kovi\'{c} Institute, Zagreb, Croatia
\item \Idef{org1298}Russian Federal Nuclear Center (VNIIEF), Sarov, Russia
\item \Idef{org1252}Russian Research Centre Kurchatov Institute, Moscow, Russia
\item \Idef{org1224}Saha Institute of Nuclear Physics, Kolkata, India
\item \Idef{org1130}School of Physics and Astronomy, University of Birmingham, Birmingham, United Kingdom
\item \Idef{org1338}Secci\'{o}n F\'{\i}sica, Departamento de Ciencias, Pontificia Universidad Cat\'{o}lica del Per\'{u}, Lima, Peru
\item \Idef{org1316}Sezione INFN, Trieste, Italy
\item \Idef{org1271}Sezione INFN, Padova, Italy
\item \Idef{org1313}Sezione INFN, Turin, Italy
\item \Idef{org1286}Sezione INFN, Rome, Italy
\item \Idef{org1146}Sezione INFN, Cagliari, Italy
\item \Idef{org1133}Sezione INFN, Bologna, Italy
\item \Idef{org1115}Sezione INFN, Bari, Italy
\item \Idef{org1155}Sezione INFN, Catania, Italy
\item \Idef{org1322}Soltan Institute for Nuclear Studies, Warsaw, Poland
\item \Idef{org36377}Nuclear Physics Group, STFC Daresbury Laboratory, Daresbury, United Kingdom
\item \Idef{org1258}SUBATECH, Ecole des Mines de Nantes, Universit\'{e} de Nantes, CNRS-IN2P3, Nantes, France
\item \Idef{org1304}Technical University of Split FESB, Split, Croatia
\item \Idef{org1168}The Henryk Niewodniczanski Institute of Nuclear Physics, Polish Academy of Sciences, Cracow, Poland
\item \Idef{org17361}The University of Texas at Austin, Physics Department, Austin, TX, United States
\item \Idef{org1173}Universidad Aut\'{o}noma de Sinaloa, Culiac\'{a}n, Mexico
\item \Idef{org1296}Universidade de S\~{a}o Paulo (USP), S\~{a}o Paulo, Brazil
\item \Idef{org1149}Universidade Estadual de Campinas (UNICAMP), Campinas, Brazil
\item \Idef{org1239}Universit\'{e} de Lyon, Universit\'{e} Lyon 1, CNRS/IN2P3, IPN-Lyon, Villeurbanne, France
\item \Idef{org1205}University of Houston, Houston, Texas, United States
\item \Idef{org20371}University of Technology and Austrian Academy of Sciences, Vienna, Austria
\item \Idef{org1222}University of Tennessee, Knoxville, Tennessee, United States
\item \Idef{org1310}University of Tokyo, Tokyo, Japan
\item \Idef{org1318}University of Tsukuba, Tsukuba, Japan
\item \Idef{org21360}Eberhard Karls Universit\"{a}t T\"{u}bingen, T\"{u}bingen, Germany
\item \Idef{org1225}Variable Energy Cyclotron Centre, Kolkata, India
\item \Idef{org1306}V.~Fock Institute for Physics, St. Petersburg State University, St. Petersburg, Russia
\item \Idef{org1323}Warsaw University of Technology, Warsaw, Poland
\item \Idef{org1179}Wayne State University, Detroit, Michigan, United States
\item \Idef{org1260}Yale University, New Haven, Connecticut, United States
\item \Idef{org1332}Yerevan Physics Institute, Yerevan, Armenia
\item \Idef{org15649}Yildiz Technical University, Istanbul, Turkey
\item \Idef{org1301}Yonsei University, Seoul, South Korea
\item \Idef{org1327}Zentrum f\"{u}r Technologietransfer und Telekommunikation (ZTT), Fachhochschule Worms, Worms, Germany
\end{Authlist}
\endgroup

\end{document}